# Frequency conversion of structured light

Fabian Steinlechner,[1,2] Nathaniel Hermosa,[1,3] Valerio Pruneri[1,4], and Juan P. Torres[1,5,*]

[1]ICFO—Institut de Ciencies Fotoniques, 08860 Castelldefels (Barcelona), Spain
[2]IQOQI—Institute for Quantum Optics and Quantum Information, Austrian Academy of Sciences, Boltzmanngasse 3, 1090 Wien, Austria
[3]National Institute of Physics, University of the Philippines Diliman
[4]ICREA—Institució Catalana de Recerca i Estudis Avançats, 08010 Barcelona, Spain
[5]Department of Signal Theory and Communications, Polytechnic University of Catalonia, Jordi Girona 1-3, 08034 Barcelona, Spain
*Corresponding author: fabian.steinlechner@univie.ac.at

**Abstract**

We demonstrate the coherent frequency conversion of structured light, optical beams in which the phase varies in each point of the transverse plane, from the near infrared (803nm) to the visible (527nm). The frequency conversion process makes use of sum-frequency generation in a periodically poled lithium niobate (ppLN) crystal with the help of a 1540-nm Gaussian pump beam. We perform far-field intensity measurements of the frequency-converted field, and verify the sought-after transformation of the characteristic intensity and phase profiles for various input modes. The coherence of the frequency-conversion process is confirmed using a mode-projection technique with a phase mask and a single-mode fiber. The presented results could be of great relevance to novel applications in high-resolution microscopy and quantum information processing.

## Introduction

The efficient generation, manipulation and detection of structured light are of great importance in numerous fields of research and technology, such as communications, quantum theory, spectroscopy, and nanoscopic imaging. In particular, beams with orbital angular momentum (OAM) [1–3] have been used for the demonstration of increased channel capacity in optical communications [4], as optical tweezers [5], in high-resolution nanoscopy [6], in measurements with increased angular resolution [7], and as a high-dimensional information carrier [8,9] for diverse applications in quantum information science.

The operational wavelength in such experiments is usually determined by the availability of light sources, the absorptivity or reflectivity of material probed, and detector sensitivity. Optical frequency conversion offers an elegant and practical way of overcoming trade-offs, which are associated with the use of a single operational wavelength and the existence of conflicting wavelength-dependent efficiencies of different components in an experiment. As an example, single-photon frequency up-conversion [10–12] from the infrared offers a convenient way for extending the working wavelength range of efficient off-the-shelf single-photon detectors and low-noise CCDs.

An ideal frequency converter transfers light to a target wavelength without distorting the information present in the field, this being a particular polarization state [13,14], temporal characteristics [15] or the spatial mode with its characteristic intensity and phase profile. To this end parametric processes in nonlinear materials [16] offer several well-established methods for converting the frequency of light, such as sum-frequency generation (SFG), second-harmonic generation (SHG), difference frequency generation (DFG), and four-wave mixing (FWM).

A key accomplishment in the field of nonlinear optics has been the frequency conversion [17–24] and amplification [25–31] of electromagnetic fields carrying spatial intensity information. The frequency conversion of spatial modes with a well-defined OAM value $l$ has been successfully demonstrated for SHG [32–34], SFG [35,36], and four-wave mixing in atomic vapor [37]. Nonlinear frequency conversion of beams in a superposition of different $l$ has also recently been demonstrated for SHG [38]. With regard to the frequency conversion via SFG, an open issue that still remains to be addressed experimentally is the coherence of the conversion process, i.e. whether it is possible to coherently frequency-convert a superposition of modes with different values of OAM. This question can be addressed by frequency-converting Hermite-Gauss (HG) beams, which can be represented as a superposition of Laguerre-Gauss ($LG^l$) beams with different OAM content $l$ [39].

Here, we report on the frequency conversion of HG spatial modes, i.e. combinations of $LG^l$, from the near infrared (803nm) to the visible spectral range (527nm). The motivation is two-fold: on one hand this demonstrates the suitability of frequency-converting OAM superposition states for applications in quantum information processing; and on the other, it shows how spatial information stored in the phase of a structured light beam can be converted to other wavelengths, which may open the door to novel applications in phase-sensitive imaging and spectroscopy.

The optical frequency converter presented here is based on SFG with a strong Gaussian pump beam. In order to efficiently frequency up-convert the phase information stored in the wave front with minimum aberrations, we make use of two key technological developments. First, the use of periodically poled nonlinear crystals, which allow the interaction to be tailored to a non-critical phase-matching configuration without spatial walk-off, while at the same time using the largest nonlinear coefficient of the material. Secondly, the availability of powerful off-the-shelf fibered pump sources around 1550nm based on erbium-doped fiber amplifiers. In the following we show that the 527-nm sum-frequency field exhibits the expected characteristic far-field intensity profile of the 803-nm input beams, and verify the preservation of coherence between different OAM values using a mode-projection technique with a phase mask and a single-mode fiber.

These results could be of great relevance for applications in quantum information processing and metrology.

## Theoretical Background

In order to get an intuitive understanding of SFG with structured light, let us briefly discuss the expected spatial distribution of the output field using standard quantum optics formalism [40]. We assume that the pump is a monochromatic coherent field, and that the signal beam is a single-photon, or weak coherent state with a structured wave front. The SFG output mode is initially in its vacuum state. The pump and the weak signal beam, each with frequencies $\omega_p$ and $\omega_s$, propagate along a principal crystallographic axis of a periodically poled nonlinear crystal of length $L$ and nonlinear coefficient $\chi^{(2)}$. The pump beam is a Gaussian beam, which in the transverse wave vector domain $\mathbf{q} = (q_x, q_y)$ writes $u_p(\mathbf{q}) = w/\sqrt{2\pi} \exp(-w^2|\mathbf{q}|^2/4)$, where $w$ is the beam waist at the center of the nonlinear crystal. The signal field's transverse wave vector distribution is $u_s(\mathbf{q})$. Due to energy and momentum conservation, the conditions $\omega_{SFG} = \omega_s + \omega_p$, and $k(\omega_p) - k(\omega_s) - k(\omega_{SFG}) - 2\pi/\Lambda = 0$ must hold, where $k(\omega)$ denotes the respective longitudinal wave vector for $\mathbf{q} = 0$, and $\Lambda$ is the poling period of the nonlinear crystal. Under these conditions the state $|\phi\rangle$ of the sum-frequency field can be expressed as:

$$|\phi\rangle = \int d\mathbf{q}\, u_{SFG}(\mathbf{q})\hat{a}^+(\mathbf{q})|vac\rangle \qquad (1)$$

where $|vac\rangle$ denotes the initial vacuum state of the sum-frequency mode, and $\hat{a}^+(\mathbf{q})$ is the creation operator for a photon with transverse momentum $q$. The spatial mode $u_{SFG}(\mathbf{q})$ of the second-harmonic wave reads:

$$u_{SFG}(\mathbf{q}) \propto \int d\mathbf{p}\, u_s(\mathbf{p}) u_p(\mathbf{q}-\mathbf{p})\, \text{sinc}(\Delta k L/2) \qquad (2)$$

with the longitudinal phase-mismatch function $\Delta k = |\mathbf{p}|^2/2k_s + |\mathbf{q}-\mathbf{p}|^2/2k_p - |\mathbf{q}|^2/2k_{SFG}$. Due to symmetry, the total OAM is conserved for collinear SFG ($l_p + l_s = l_{SFG}$). Since the Gaussian pump beam carries no OAM ($l_p = 0$), the OAM content of the signal beam is transferred to the SFG field ($l_{SFG} = l_s$). However, the spatial mode generated in the SFG process may be distorted with respect to the input mode as a result of spatial filtering due to: a) the limited spatial extent of the pump mode and b) the spatial bandwidth of the phase-matching function [41]. If the length of the nonlinear crystal is short compared to the diffraction length of the pump and signal beams, the expression for the spatial mode of the SFG simplifies to:

$$u_{SFG}(\mathbf{q}) \propto \int d\mathbf{p}\, u_s(\mathbf{q})\delta(\mathbf{q}-\mathbf{p}) = u_s(\mathbf{q}) \qquad (3)$$

Therefore, one obtains a perfect frequency-converted replica of the signal field. A detailed analysis of the effects of spatial filtering is beyond the scope of this article [42,27]. In the following we restrict our considerations to the conservation of the phase structure of superposition of modes with different values of OAM.

## Experimental Setup

The experimental setup depicted in Fig. 1 consists of three stages. In the first, the input beam is manipulated in order to obtain the desired structured wave front. In the second stage, the frequency of the incident NIR field is converted to a visible wavelength using SFG in a nonlinear crystal. Finally, the spatial characteristics of the sum-frequency mode are analyzed via far-field intensity measurements, and a coherent mode-projection technique. Each stage is discussed in more detail in the following.

### Encoding the phase structure

Laser light from a fiber-coupled 803-nm continuous-wave laser diode is collimated and impinges on a spatial light modulator (SLM). The first-order diffraction of the SLM is modulated in phase and amplitude [43], in order to generate either $LG^l_{p=0}$ modes (radial mode index p=0), superpositions of $LG^l_0$ and $LG^{-l}_0$, or $HG_{nm}$ modes, where $m$ and $n$ are the mode indices in the x and y directions. A spatial filter after the SLM removes the zero-th order diffraction from the SLM. The spatially modulated 803-nm input beam is then directed to the frequency conversion stage. The expected far-field intensity profiles were verified for each input mode by inserting a CCD located in the Fourier plane of 2-f imaging system after the spatial filter (not depicted in Fig. 1).

### Frequency conversion

The frequency conversion setup uses SFG in a 10-mm-long Magnesium-doped ppLN crystal (Covesion Ltd.) with a poling period of 7.8 μm. The crystal is maintained at a temperature of approx. 85°C, for SFG with quasi-phase matched center wavelengths of $\lambda_p$=1540nm, $\lambda_s$=803nm, and $\lambda_{SFG}$=527nm. A fiber-coupled 1540-nm continuous-wave laser diode is amplified to a power of approx. 60mW via an Erbium-doped fiber amplifier, and the output beam of the single-mode fiber is collimated and magnified. The pump beam is combined with the input beam via a dichroic mirror which is highly reflective at 803nm and highly non-reflective at 1540nm. For the mixing process to occur in the Fourier plane, both beams are focused to the center of the nonlinear crystal using an achromatic $f$=100-mm lens. The pump and signal beams propagate along the Mg:ppLN crystal's optical z-axis (non-critical configuration). This avoids the introduction of wave-front distortions due to spatial walk-off, and allows using the large nonlinear $d_{33}$ coefficient of LN, which leads to overall higher conversion efficiency. The SFG beam is collected using an $f$=200-mm lens, and the pump light is guided to a beam dump using a dichroic mirror which reflects the 1540-nm and transmits the 527-nm light. The remaining pump light, and the 803-nm signal, are then blocked using a short-pass filter and a narrowband interference filter with a passband of approximately 10nm, centered around 532-nm.

### Detection

A flip mirror directs the 527-nm sum-frequency beam to a module for either a phase- or an intensity measurement. The intensity measurements are performed in the far field using CCD located at the Fourier plane of a 2-f imaging system. The phase structure of the outgoing beam is analyzed by projecting onto an appropriately tailored spatial mode [44–46]. The overlap with the $LG_0^0$ mode, which has a flat phase front, can be evaluated by directly imaging the SFG output onto the input facet of a 532-nm single-mode fiber. For the projection onto a $HG_{10}$ mode, an additional phase mask is inserted before the single-mode fiber. The π phase flip required for the projection onto a $HG_{10}$ mode is obtained by angle-tuning a transparent microscope slide of appropriate thickness which is inserted half-way into the beam path. The overlap of both the input mode and the mode generated by the nonlinear crystal with the projection mode is evaluated by monitoring the power coupled into the single-mode fiber with an optical power meter. For the mode projection of the input beam, an additional phase mask and an 810-nm single-mode fiber (not depicted in Fig. 1) were inserted into the beam path before the nonlinear crystal.

## Results

One of our main results can be visualized using a Poincaré sphere equivalent for OAM [47,48] (Fig. 2). The polar points on the sphere correspond to $LG_0^l$ modes with $l = \mp 1$. The points on the equatorial plane correspond to OAM superposition states $HG_{10}(\theta) \propto e^{i\theta} LG_0^1 + e^{-i\theta} LG_0^{-1}$. These superpositions correspond to a $HG_{10}$ mode rotated by an angle θ [$HG_{10}(\pi/2) = HG_{01}(0)$].

By changing the relative phase and magnitude of the OAM components one can access states distributed over the entire Poincaré sphere. For all phase profiles applied to the input beam, the 527-nm frequency-converted SFG field exhibits the same characteristic far-field intensity profile as the 803-nm input mode (inset Fig. 2). The fact that the angular orientation of the sum-frequency mode is the same as that of the signal beam demonstrates that relative phase of the OAM states is unaffected by the frequency-conversion process.

For 60mW of pump power (47 mW impinging on the nonlinear crystal) we observed power-conversion efficiencies (P$_{SHG}$/P$_{signal}$) of approx. 0.11% for the $LG_0^0$ mode (not depicted in Fig. 2), 0.09% for the $LG_0^{\pm 1}$ modes, and 0.08% for the $HG_{01}$ and $HG_{10}$ modes. Achieving higher conversion efficiencies is the subject of further research, and we believe that significant improvements can be expected by placing the nonlinear crystal inside an optical cavity, or by optimization of alignment and confocal parameters of the pump- and signal beams. In this proof-of-concept experiment, however, our primary goal was to demonstrate that the phase coherence is maintained in the frequency-conversion process. While this claim is already supported by the results discussed in the preceding paragraph, we also performed phase-dependent measurements using a coherent mode projection technique.

As a first step we evaluated the overlap of both the input beam, and SFG with an $LG_0^0$ projection mode. Fig. 3(a) depicts the normalized power coupled into the $LG_0^0$ mode of a single-mode fiber with various phase profiles applied to the input beam via the SLM. As expected, only the flat phase front of the $LG_0^0$ input beam leads to efficient coupling into the projection fiber. Conversely, the $HG_{01}$, $HG_{10}$, and $LG_0^{\pm 1}$ modes show negligible overlap with the $LG_0^0$ projection mode. This behavior was observed for both the input, and output modes. Similar behavior was observed when a phase mask with a π phase step was added before the projection fiber. This approximately corresponds to the projection onto the anti-symmetric $HG_{10}$ mode (Fig. 3(b)). With the phase mask in place, the $LG_0^0$ mode of both input and output beams resulted in a negligible amount of power coupling into the single mode fiber. In this case, maximal overlap was observed for a $HG_{10}$ mode, for which the phase mask effectively flattens the wave front. The $HG_{01}$ mode, on the other hand, has negligible overlap with this projection mode, as a result of the orthogonality of the HG modes. In the case of $LG_0^{\pm 1}$ modes approximately 50% of the power couples into the single-mode fiber for both input and output beams. This behavior becomes clear when one when one considers the $LG_0^{\pm 1}$ as an equal superposition of $HG_{10}$ and $HG_{01}$ modes. These results lead us to the conclusion that the phase relationship between different OAM modes is conserved, thus demonstrating the suitability of the scheme for coherently converting OAM superposition states for applications in quantum information processing.

In order to extend these results to higher OAM values, and furthermore assess the feasibility of applying the technique, for example, in the coherent imaging of more complex phase structures, we also frequency-converted several higher-order spatial modes.

The spatial intensity profiles (Fig. 4) of the 537-nm mode showed a high degree of similarity with the 803-nm input mode. However, for higher mode numbers aberrations became apparent. This is the result of crystal inhomogeneity, beam clipping effects at the limited aperture of the nonlinear crystal, as well as spatial filtering effects due to the dimensions of waist of the pump beam. This effect could, however, be mitigated using both a larger crystal aperture, as well as a larger beam waist. Determining the pump focus parameters which lead to the highest conversion-efficiency for a particular higher-order spatial mode, as well as the transfer function for more general distributions of transverse momenta should be the subject of further research.

## Conclusions

We have demonstrated coherent frequency conversion of Laguerre-Gauss and Hermite-Gauss modes from the near infrared to the visible via the process of sum-frequency mixing in a periodically poled Mg:LN crystal. Our results demonstrate conclusively that the relative phase between Laguerre-Gauss modes with different OAM content is maintained in the frequency-conversion process. This establishes the feasibility of frequency converting quantum information encoded in OAM superposition states.

More generally, this can be seen as a method for trans-frequency imaging of a phase-only object, and we believe that more complex phase-structured optical beams may be converted using similar techniques. This could be of great relevance to applications in microscopy, since it allows acquiring phase information about an object from light at a wavelength different from the one used to illuminate the object [49] .

However, further improvements are required in order to obtain higher conversion efficiencies as required for wide-scale practical applicability in different scenarios. Nevertheless, our results indicate the experimental feasibility of proof-of-concept experiments involving the coherent frequency conversion of multidimensional quantum information encoded in the spatial degree of freedom of (entangled) single photons, which could be of great significance for future applications in quantum information processing and sensing.

The techniques employed in this article could also be applied to optical parametric amplification, thus enabling the generation of desired spatial modes with high power from a weak seed field.

**Funding.** JPT and VP acknowledge support from ICREA (Generalitat de Catalunya). The work has received financial support from the Spanish Ministry of Economy and Competitiveness (MINECO), the Fondo Europeo de Desarrollo Regional (FEDER) through grant TEC2013-46168-R. FS acknowledges financial support via MINECO FPI fellowship.

**Acknowledgements.** FS acknowledges the valuable support of Adam Valles, as well as helpful discussions with Roberto de León-Montiel and Carmelo Rosales Guzmán.

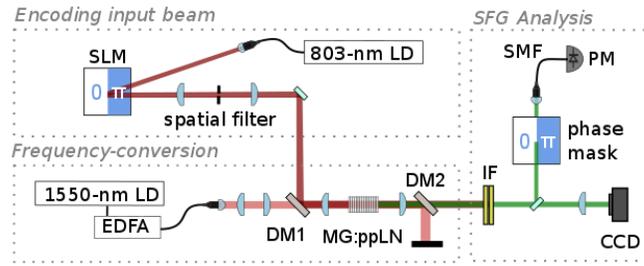

Fig. 1. Experimental setup. The phase front of light from an 803-nm laser diode (LD) is modulated using a spatial light modulator (SLM). The 803-nm light (dark red) is reflected from a dichroic mirror (DM1) and frequency conversion is accomplished in a Mg:ppLN crystal pumped with a 1540-nm LD (light red). Another dichroic mirror (DM2) blocks the pump light, and transmits the 803-nm beam and the 527-nm sum-frequency beam (green). The 803-nm light is blocked with an interference filter (IF). A flip mirror is used to select between analysis of the phase structure or the intensity profile of the sum-frequency field. The phase-structure is analyzed using a phase mask and a single-mode fiber (SMF). The power coupled into the single-mode fiber is monitored using a power meter (PM). The intensity profile of the sum-frequency field was analyzed using a CCD.

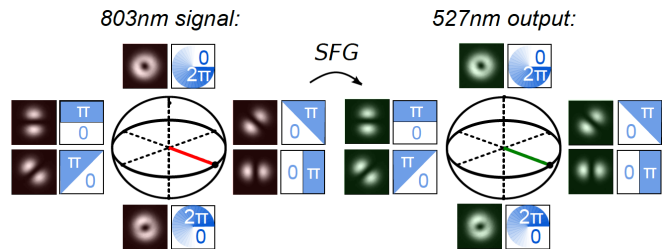

Fig. 2. Measured far-field intensity profiles of the 803-nm input beam (left, red) and the frequency-converted 527-nm output beam (right, green) for several phase profiles applied via the SLM in the 803-nm input beam. The relative phase of the $\text{LG}_0^{\pm 1}$ modes determines the angular orientation of the $\text{HG}_{10}(\theta) \propto e^{i\theta}\text{LG}_0^1 + e^{-i\theta}\text{LG}_0^{-1}$ modes on the equator of the Poincaré sphere for OAM.

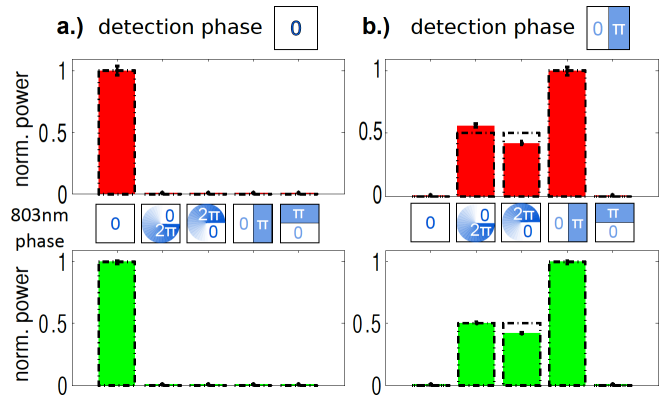

Fig. 3. Demonstration of the transfer of the phase structure to the SFG output beam. The bar plots depict the normalized power measured after mode projection with a uniform-phase mask (3(a)) or a phase-flip mask (3(b)) for various input modes (red, top) and SFG output modes (green, bottom). The error bars due to power fluctuations are small in comparison to the systematic errors due to imperfect alignment of the SLM and projection mode. The dashed line (black) depicts the expected theoretical behavior.

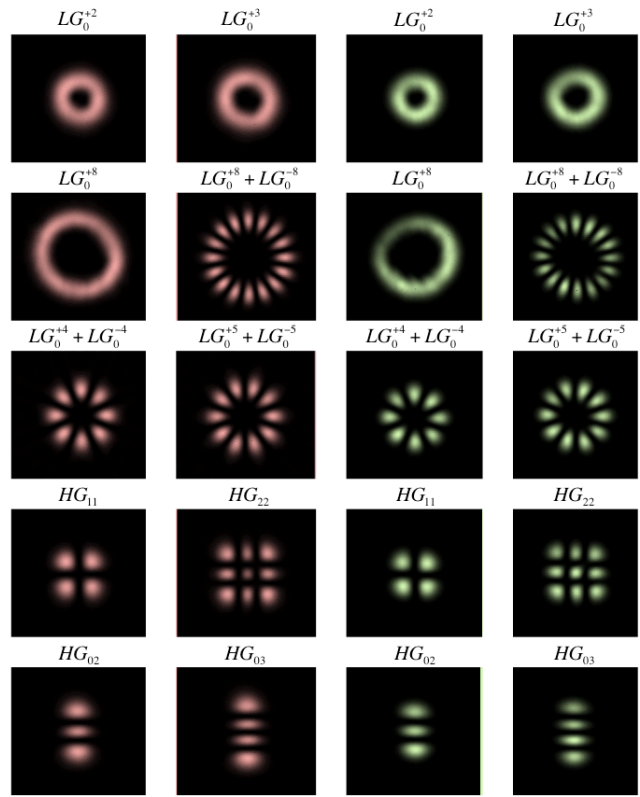

Fig. 4. Measured far field intensity profiles of various higher-order spatial modes for the input signal beam (left) and the frequency-converted beam (right). The fidelity of the frequency-converted spatial intensity patterns demonstrates the feasibility of coherently frequency converting complex coherent spatial patterns, as well as higher-dimensional quantum systems.